\documentclass[11pt]{article}
\usepackage{jheppub} 
\usepackage[utf8]{inputenc}
\usepackage[T1]{fontenc}
\usepackage[english]{babel}
\usepackage{amsmath}
\usepackage{amsfonts}
\usepackage{amssymb}
\usepackage{mathtools}
\usepackage{enumitem}
\usepackage{graphicx}
\usepackage{bbm}
\usepackage{color}
\usepackage{xspace}
\usepackage{url}

\newcommand{\cE}{{\mathcal E}}

\newcommand{\cM}{{\mathcal M}}

\newcommand{\cO}{{\mathcal O}}

\title{On the non-commutativity of geometric observables in different Lorentz frames}

\author[1]{Mehdi Assanioussi,}
\author[1,2]{Jerzy Kowalski-Glikman,}
\author[1]{Ilkka M\"akinen,}
\author[1]{Ludovic Varrin}
\affiliation[1]{National Centre for Nuclear Research, Pasteura 7, 02-093 Warsaw, Poland}
\affiliation[2]{Faculty of Physics and Astronomy, University of Wroclaw, Pl. Maksa Borna 9, 50-204 Wroclaw}

\emailAdd{mehdi.assanioussi@ncbj.gov.pl}
\emailAdd{jerzy.kowalski-glikman@uwr.edu.pl}
\emailAdd{ilkka.makinen@ncbj.gov.pl}
\emailAdd{ludovic.varrin@ncbj.gov.pl}

\begin{document}
\abstract{
Our aim is to establish whether geometric observables, such as length, area or volume of a physical object, viewed by different observers Poisson commute or not. To illustrate this, we compute the Poisson bracket of two lengths associated to a rigid rod and measured by two different geodesic (inertial) observers, one of which is at rest while the other is moving with respect to the rod. Our calculation shows that geometric observables measured by different observers generically do not Poisson commute, not even in Minkowski spacetime. This non-trivial result provides interesting insights into questions related to the presence of a fundamental scale in the context of quantum gravity.
}

\maketitle

\newpage

\setcounter{page}{1}

\section{Introduction}

Is Lorentz (or more generally Poincaré) symmetry preserved once quantum gravity effects are taken into account? This question has remained open for at least a quarter of a century \cite{Amelino-Camelia:1999hpv}. It is a central issue, given that essentially all quantum gravity research programs agree on the existence of a minimal observable length, as emphasized by Mead \cite{Mead:1964zz}, presumably of the order of the Planck length $\ell_P$, which suggests a discrete structure at the Planck scale. In Mead's approach, but also in other similar constructions reviewed in \cite{Hossenfelder:2012jw}, the minimal length arises because achieving better and better resolution requires the use of probes of increasingly high energy, so much that eventually the entire observed region collapses into a black hole. If one interprets this as a signal of a fundamental discreteness of spacetime, one immediately encounters tension with Lorentz/Poincaré symmetry, although there are proposals that may resolve this tension \cite{Freidel:2016pls}. It is worth noting that the question of Lorentz symmetry at the Planck scale seems to have now entered the domain of quantum gravity phenomenology, and it is not impossible that we will obtain reliable observational evidence addressing this issue sooner rather than later \cite{Addazi:2021xuf}.

The fate of Lorentz symmetry at the Planck scale has seemed particularly intriguing in the context of Loop Quantum Gravity \cite{Rovelli:2004tv, Thiemann:2007pyv}, which predicts a discrete spectrum for length, area, and volume operators \cite{Rovelli:1994ge, Ashtekar:1996eg, Thiemann:1996at, Ashtekar:1997fb, Bianchi:2008es} (see however \cite{Dittrich:2007th, Rovelli:2007ep} for discussion). A common argument \cite{Amelino-Camelia:2002cfs} states that because quantum gravity should reproduce special relativity in regimes where energies are far below the Planck energy scale and spacetime curvature is negligible, one must ask for which Lorentz observer the minimal length remains minimal. If one observer measures a minimal length, a boosted observer would, in principle, see that length contracted. Usually, two proposals are considered to resolve this conundrum. If for some reason the Lorentz symmetry is broken at the Planck scale, and a preferred observer exists, minimal length is defined in this observer's frame. Alternatively, one can deform the Lorentz (Poincar\'e) symmetry in such a way that all inertial observers measure the same value of the minimal length \cite{Amelino-Camelia:2000cpa,Amelino-Camelia:2000stu,Magueijo:2001cr}, in other words the fundamental length (and mass) scale is left invariant by the modified, deformed Lorentz transformations.

An interesting yet different perspective to address this issue has been discussed in \cite{Rovelli:2002vp, Rovelli:2010ed}, where the authors argue that the spatial area (and similarly length or volume) operators associated with different Lorentz frames do not commute. As a result, if one observer measures an area eigenvalue in a given state, the corresponding area operator for a boosted observer necessarily acquires a large uncertainty (see also \cite{Varadarajan:2026ekv} for a more recent analysis of the interplay between discrete spectra and Lorentz boosts in the context of a quantized parameterized field theory). The question of which Lorentz observer measures the minimal area thus becomes irrelevant, since it cannot be sharply measured by two observers simultaneously. Notice that the argument here is rather different from the minimal length argument of Mead: The reasoning is purely kinematical and as such cannot involve black holes. 

In this article we argue that the non-commutativity of spatial geometric observables such as lengths, areas and volumes, is a much more general effect than suggested in \cite{Rovelli:2002vp}: It is in fact rooted in the canonical structure of gravity and is therefore of quite universal nature. The effect is present even if the gravitational field is switched off, i.e.~in Minkowski spacetime.

The plan of the article is as follows. In the next section, we provide the general setup for our investigation. In Section 3, we carefully define the simultaneity surfaces and the corresponding induced metrics, needed to define geometric observables for a given observer. Section 4 presents the main result of our work: The calculation of the Poisson bracket of length observables associated with two simultaneity surfaces corresponding to two Lorentz observers. We find that this Poisson bracket generically does not vanish, not even in Minkowski spacetime. The final Section is devoted to discussing the significance and implications of this result.

\section{Context and general considerations}

Given a spacetime $\cM$ with a metric $g_{\mu\nu}$, consider a free falling rigid rod with an observer $\cO$ attached to it at a point $p$ and with proper time $t$. The worldline of the observer $\cO$ is a geodesic passing through the point $p$ and having a certain tangent vector $u^\mu$ at that point. Then consider a second observer $\bar \cO$ with proper time $\bar t$ moving along a geodesic determined by a different tangent vector $\bar u^\mu$ at the point $p$.
We take a small spacetime neighborhood $\cE$ around the point $p$, with a characteristic size $\epsilon$, and we consider the whole world sheet $R$ of the rod for a finite time interval $\Delta t$ to be covered by $\cE$. The local setting is sufficient to establish the result regarding commutativity.
We choose the coordinates\footnote{Throughout the article, Greek letters denote spacetime indices while Latin letters denote spatial indices. The value $0$ denotes the time direction while the values $1,2,3$ denote the three spatial directions.} $x^\mu=(t,x^1,x^2,x^3)$ in $\cE$ with origin at $p$, such that $u^\mu = (1, 0, 0, 0)$, $\bar u^\mu = (1, \beta, 0, 0)$ and $g_{0\mu}=-\delta_{0\mu}$ in these coordinates. This gauge fixing does not alter significantly the results of the Poisson brackets we are interested in. These considerations guarantee that all the observables and calculations are evaluated locally in a simple form.

\begin{figure}[t]
    \centering
    \includegraphics[width=0.34\linewidth]{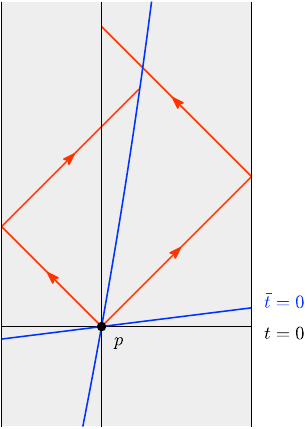}
    \caption{Depicting the world sheet $R$ of a rod (grey) and the process of measuring its length, via light signals (red) emitted at the point $p$, by two geodesic observers: One at rest with respect to the rod (black) and one in motion with respect to it (blue).}
    \label{fig:Measurement}
\end{figure}

Since our objective is to establish whether geometric observables viewed by different observers Poisson commute or not, we attempt to compute the Poisson bracket of the lengths of the rod as measured by the two aforementioned observers. The result in this case would be sufficient to answer the question at hand. But before moving to the calculations, let us briefly discuss how a measurement of length can be operationally performed. This would help us identify the proper setup to adopt in our calculations. For simplicity of this discussion, we for now assume Minkowski spacetime. If one of the two observers is to measure the length of the rod, one operational method is to use light signals to measure the distance between the two ends of the rod. Namely, consider having a pair of perfectly reflective mirrors, each placed at each end of the rod. If, at proper time $\tau=0$, the observer in question is at a point $p$ of the rod which is different from its end points, then they would need to send two light signals, in opposite directions, each towards one end of the rod. The observer would then measure the time it takes to receive the reflected signals on the mirrors. Generically, the two light signals do not take the same amount of the observer's proper time to come back, therefore the observer must make two measurements of time, one for each signal, as illustrated in figure \ref{fig:Measurement}.

In other words, to measure the length of the rod from an arbitrary point $p$, the observer in fact makes two measurements, namely measuring the distances between the observer and the two ends of the rod. Classically, such an observation is not really relevant for the interpretation of the result of the calculation of Poisson brackets, however, in a quantum theory, the operational aspect of the measurement matters. Therefore, it is appropriate to consider the setup where only one measurement is needed. This is achieved by simply considering the point $p$, where the observer initiate their measurement, to coincide with one of the ends of the rod. As we will see, this restriction manifests only in the domain of integration over the manifold and we will further comment on it later on. This whole analysis regarding the measurement of length also stands in the context of arbitrary spacetime. However, in curved spacetime and in the absence of any knowledge of the metric, the operational measurement of distances (through the measurement of propagation time and other properties of light signals e.g.~frequency) cannot be accurate and an observer can only estimate a sort of ``average'' or approximation at best of the distance via light signals. Nevertheless, in a small spacetime neighborhood where the variation of curvature is not large, the measurement can be approximated to a reliable accuracy, hence the local considerations above.

Now that our setup is defined, we can move to the derivation of the main results of the article. We begin by noting that in order to define spatial geometric observables for a given observer, we need to identify the simultaneity surface of the observer. The simultaneity surface determines the induced metric to be used to define the observables. There are various definitions of the simultaneity surface in curved spacetimes \cite{Misner:1973prb, Lachieze-Rey:2001rrt, Bolos:2002qp}, but in the present article, we consider the definition based on the spacetime light-cones which we present in the next section. Generally, the exact calculation of the simultaneity surface is highly non-trivial. However, in a local setting, we are able to use an expansion in the coordinates. In the following section, we show how to derive the equation defining the simultaneity surface for an arbitrary geodesic observer using the local expansion, then we provide the expression of the corresponding induced metric.

\section{Simultaneity surface and induced metric}

Consider an observer moving along a given world line $y^\mu(s)$, with a coordinate system chosen such that the observer's position at proper time $s=0$ corresponds to the origin $p$ of the coordinate system. The observer's surface of simultaneity in a small neighborhood around the origin can be determined by the following construction:
For a given proper time value $s>0$, take the two points $x^\mu_+ = y^\mu(s)$ and $x^\mu_- = y^\mu(-s)$ belonging to the observer's world line, respectively in the future and past of the origin, and consider the past lightcone at $x^\mu_+$ and the future lightcone at $x^\mu_-$; then the set of points determined by the intersection of the two lightcones belongs to the observer's surface of simultaneity. As the proper time value $s$ varies, the union of the intersection sets (one for each value of $s$) forms the simultaneity surface in the small neighborhood of interest.

In order to derive the simultaneity surface explicitly, it is useful to first consider the following auxiliary problem: Given a fixed point $x^\mu_0$, which nearby points $X^\mu$ are connected to $x^\mu_0$ by a null geodesic? For such a geodesic to exist, there must be a null vector $v^\mu$ (at the point $x^\mu_0$) and a value of an affine parameter $\lambda$ such that\footnote{Throughout the article, we adopt the notation where the metric tensor and its derivatives, including the Christoffel symbols $\Gamma^\mu_{\alpha\beta}$, written without a spacetime point as an argument means that they are evaluated at the origin $p$ of the coordinate system.}
\begin{equation}
    X^\mu = x^\mu_0 + v^\mu \lambda + \frac{1}{2}a^\mu \lambda^2 + {\cal O}(\lambda^3)
    \label{eq:connectivity}
\end{equation}
where $a^\mu = -\Gamma^\mu_{\alpha\beta}v^\alpha v^\beta$. Using eq.~\eqref{eq:connectivity}, we can extract the initial tangent vector in terms of the starting and ending points. We have
\begin{align}
    v^\mu &= \frac{X^\mu - x^\mu_0}{\lambda} + \frac{1}{2}\Gamma^\mu_{\alpha\beta}v^\alpha v^\beta \lambda + {\cal O}(\lambda^2)
    \label{eq:v^mu}
\end{align}
which shows that at lowest order in $\lambda$, $v^\mu = (X^\mu - x^\mu_0)/\lambda + {\cal O}(\lambda)$. Using this in eq.~\eqref{eq:v^mu}, we find
\begin{align}
    v^\mu &= \frac{1}{\lambda}\biggl(\xi^\mu + \frac{1}{2}\Gamma^\mu_{\alpha\beta}\xi^\alpha\xi^\beta\biggr) + {\cal O}(\lambda^2)
    \label{}
\end{align}
where
\begin{equation}
    \xi^\mu = X^\mu - x^\mu_0.
\end{equation}
Therefore the condition for $x^\mu_0$ and $X^\mu$ to be connected by a null geodesic (at quadratic order in $\lambda$) is that the vector $n^\mu$, whose value at $x_0$ given by
\begin{equation}
    n^\mu = \xi^\mu + \frac{1}{2}\Gamma^\mu_{\alpha\beta}\xi^\alpha\xi^\beta,
    \label{eq:n}
\end{equation}
is null at $x^\mu_0$:
\begin{equation}
    g_{\mu\nu}(x_0)n^\mu n^\nu = 0.
    \label{eq:null}
\end{equation}

Let us now apply this to the construction of the simultaneity surface for a geodesic observer, whose worldline is
\begin{equation}
    y^\mu(s) = u^\mu s + \frac{1}{2}a^\mu s^2 + \mathcal{O}(s^3)
    \label{eq:world line}
\end{equation}
where $u^\mu = (1, u^a)$ and $a^\mu = -\Gamma^\mu_{\alpha\beta}u^\alpha u^\beta$. For proper time $s_0 > 0$, the starting points for the past and future lightcones are $x^\mu_\pm = y^\mu(\pm s_0)$. Consider the point $X^\mu = \bigl(x^0, x^1, x^2, x^3\bigr)$ with the spatial coordinates $x^a$ fixed and $x^0$ unknown. For $X^\mu$ to lie on the simultaneity surface, both $x^\mu_+$ and $x^\mu_-$ must be connected to $X^\mu$ by null geodesics. Imposing the condition \eqref{eq:null} at $x^\mu_+$ and $x^\mu_-$ gives two equations, which determine the two unknowns $x^0$ and $s_0$.

At lowest order in $s_0$, we may ignore the quadratic terms in \eqref{eq:n} and \eqref{eq:world line} -- note that $\xi^\mu = {\cal O}(s_0)$ -- and use $g_{\mu\nu} = g_{\mu\nu}(0)$ instead of $g_{\mu\nu}(x_\pm)$ in \eqref{eq:null}. Then we have the conditions
\begin{equation}
    g_{\mu\nu}\xi^\mu_+\xi^\nu_+ = 0 \qquad \text{and} \qquad g_{\mu\nu}\xi^\mu_-\xi^\nu_- = 0
    \label{eq:lowest}
\end{equation}
where
\begin{equation}
    \xi^\mu_\pm = X^\mu - x^\mu_\pm = X^\mu \mp u^\mu s_0.
\end{equation}
Since $s_0 \neq 0$, the solution for $x^0$ and $s_0$ is given by
\begin{equation}
    x^0 = g_{ab}u^a x^b, \qquad
    s_0 = \sqrt{\frac{g_{ab}x^a x^b - \bigl(g_{ab}u^a x^b\bigr)^2}{1 - g_{ab}u^a u^b}}.
    \label{x0first_order}
\end{equation}
Going to the next order in $s_0$, the solution found above receives a correction:
\begin{equation}
    x^0 \to x^0 + \delta x^0, \qquad s_0 \to s_0 + \delta s_0
    \label{}
\end{equation}
where $\delta x^0$ and $\delta s_0$ are of order $s_0^2$. Hence the starting points for the lightcones will now be
\begin{equation}
    x^\mu_\pm \to y^\mu\bigl(\pm(s_0 + \delta s_0)\bigr) = x^\mu_\pm + \delta x^\mu_\pm.
    \label{}
\end{equation}
with $\delta x^\mu_\pm = \pm u^\mu\,\delta s_0 + \frac{1}{2}s_0^2 a^\mu$.
Accordingly, $\xi^\mu_\pm$ are replaced by $\xi^\mu_\pm + \delta\xi^\mu_\pm$, where
\begin{equation}
    \delta\xi^\mu_\pm = \delta X^\mu - \delta x^\mu_\pm = \delta X^\mu \mp u^\mu\,\delta s_0 - \frac{s_0^2}{2}a^\mu
    \label{}
\end{equation}
and $\delta X^\mu = (\delta x^0, 0, 0, 0)$. Then each vector $n^\mu_\pm$ is
\begin{equation}
    n^\mu_\pm = \xi^\mu_\pm + \delta\xi^\mu_\pm + \frac{1}{2}\Gamma^\mu_{\alpha\beta}\xi^\alpha_\pm\xi^\beta_\pm + {\cal O}(s_0^3).
    \label{}
\end{equation}
The corrections $\delta x^0$ and $\delta s_0$ are now determined by imposing the conditions
\begin{equation}
    g_{\mu\nu}(x_\pm)n^\mu_\pm n^\nu_\pm = 0
    \label{}
\end{equation}
at next-to-leading order in $s_0$. Collecting the terms of order $s_0^3$, we have the equations
\begin{align}
    s_0u^\lambda\partial_\lambda g_{\mu\nu}\xi^\mu_+\xi^\nu_+ + 2g_{\mu\nu}\xi^\mu_+\,\delta\xi^\nu_+ + g_{\mu\nu}\xi^\mu_+\Gamma^\nu_{\alpha\beta}\xi^\alpha_+\xi^\beta_+ &= 0 \\[2ex]
    -s_0u^\lambda\partial_\lambda g_{\mu\nu}\xi^\mu_-\xi^\nu_- + 2g_{\mu\nu}\xi^\mu_-\,\delta\xi^\nu_- + g_{\mu\nu}\xi^\mu_-\Gamma^\nu_{\alpha\beta}\xi^\alpha_-\xi^\beta_- &= 0,
\end{align}
which can be rearranged into the form
\begin{equation}
    \xi^0_\pm\,\delta x^0 \pm g(u, \xi_\pm)\,\delta s_0 = F_\pm(\xi_\pm)
    \label{eq:eq+-}
\end{equation}
where 
\begin{equation}
    F_\pm(\xi) = -\frac{s_0^2}{2}g(a, \xi) \pm \frac{s_0}{2}\partial g(u, \xi, \xi) + \frac{1}{4}\partial g(\xi, \xi, \xi)
    \label{eq:F}
\end{equation}
with the notation $g(u, v) = g_{\mu\nu}u^\mu v^\nu$ and $\partial g(u, v, w) = \partial_\lambda g_{\mu\nu}u^\lambda v^\mu w^\nu$. Solving eqs.~\eqref{eq:eq+-} for $\delta x^0$ now yields
\begin{equation}
    \delta x^0 = \frac{g(u, \xi_+)F_-(\xi_-) + g(u, \xi_-)F_+(\xi_+)}{\xi_+^0 g(u, \xi_-) + \xi_-^0 g(u, \xi_+)} = \frac{1}{2s_0}\Bigl(F_-(\xi_-) - F_+(\xi_+)\Bigr)
    \label{eq:dx0[F]}
\end{equation}
where the second step follows after observing from \eqref{x0first_order} that $g(u, X) = 0$. Next, we rewrite $F_\pm(\xi_\pm)$ by inserting $\xi^\mu_\pm = X^\mu \mp u^\mu s_0$ into eq.~\eqref{eq:F}, finding
\begin{equation}
    F_\pm(\xi_\pm) = \frac{1}{4}\partial g(X, X, X) \pm \frac{s_0}{4}\Bigl(\partial g(u, X, X) - 2\partial g(X, u, X)\Bigr).
    \label{}
\end{equation}
Using this in eq.~\eqref{eq:dx0[F]} and performing some straightforward algebra, we obtain
\begin{align}
    \delta x^0 = -\frac{1}{4}\dot g_{ab}x^a x^b + \frac{1}{2}u^c\Gamma_{cab}x^ax^b + \frac{1}{2}x^0\dot g_{ab}u^a x^b
    \label{eq:dx0[x]}
\end{align}
where $\Gamma_{abc} = \frac{1}{2}\bigl(\partial_b g_{ac} + \partial_c g_{ab} - \partial_a g_{bc}\bigr)$.
Combining eq.~\eqref{eq:dx0[x]} with the first-order result $x^0 = g_{ab}u^a x^b$, we have the complete solution at quadratic order in the coordinates:
\begin{equation}
    x^0 = g_{ab}u^a x^b -\frac{1}{4}\dot g_{ab}x^a x^b + \frac{1}{2}u^c\Gamma_{cab}x^a x^b + \frac{1}{2}\bigl(g_{ab}u^a x^b\bigr)\bigl(\dot g_{ab}u^a x^b\bigr).
    \label{Sim.Surf.}
\end{equation}
Equation \eqref{Sim.Surf.} is the defining equation of the simultaneity surface of an observer with a world line characterized by the tangent vector $u^\mu$, in the approximation given by the expansion in terms of the coordinates $x^\mu$.

It follows that the induced metric on the surface defined by \eqref{Sim.Surf.} is given by
\begin{align}
    h_{ab}(x) &= g_{\mu\nu}(x) \frac{\partial x^\mu}{\partial x^a} \frac{\partial x^\nu}{\partial x^b} = g_{ab}(x) - u_a u_b - u_a\bigl(-\dot g_{bc} + 2 u^d\Gamma_{dbc} + 2 \dot g_{d(c} u_{b)} u^d \bigr)x^c 
    \label{Ind.metric}
\end{align}
with $u_a = g_{ab} u^b$. We remind the reader that the notation of tensors without spacetime points arguments means that these tensors are evaluated at the origin $p$.

An interesting observation is that the results in \eqref{Sim.Surf.} and \eqref{Ind.metric} show that if one is to consider the diffeomorphism which would map the induced metric of one observer to the induced metric of a boosted observer, identified by the tangent vectors of their respective world lines at their intersection, such diffeomorphism would be phase space dependent. This means that one can define observers boosted with respect to each other in a local and background independent fashion, using simply the tangent vectors at the intersection of the world lines, and that naturally leads to identifying a phase space dependent diffeomorphism which maps one observer's proper time and simultaneity surface to the other's, in any spacetime.

\section{Lengths in different frames and their Poisson brackets}
\label{sec:lengths}

For simplicity, we choose the world sheet $R$ of the rod to be defined as $x^\mu=(t,x,0,0)$, with $0 \leqslant x \leqslant \epsilon$. 
At the instant $t=0=\bar t$, the world lines of the two observers $\cO$ and $\bar \cO$ intersect at the point $p$ with $x^\mu(p)=(0,0,0,0)$ and tangent vectors $u^\mu = (1,0,0,0)$ and $\bar u^\mu = (1,\beta,0,0)$ respectively. Using eq.~\eqref{Ind.metric}, it follows that the induced metrics $h$ and $\bar h$ on the rod for $\cO$ and $\bar \cO$ respectively are given by
\begin{align}
    h(x) &= g_{11}(0,x,0,0) \\[2ex]
    \bar h(x) &= g_{11}\bigl(t(x),x,0,0\bigr) - \beta^2 g_{11}^2 - \beta g_{11} \left(-\dot g_{11} + 2 \beta \Gamma_{111} + 2 \beta^2 g_{11} \dot g_{11} \right) x
    \label{rod.metric}
\end{align}
where $t(x)$ is given by equation \eqref{Sim.Surf.} for $u^a = (\beta,0,0)$ and $x^a=(x,0,0)$.
Therefore, the lengths $L$ and $\bar L$ of the rod at the instants $t=0=\bar t$ are given by
\begin{equation}
    L = \int_{R(t=0)} dx \sqrt{h(x)}\ , \qquad \bar L = \int_{R(\bar t=0)} dx \sqrt{\bar h(x)}.
    \label{lengths}
\end{equation}
Note that in Minkowski spacetime, the metric is constant and the two simultaneity surfaces are related by a standard Lorentz transformation characterized by the relative velocity $\beta$. From \eqref{lengths}, one can see that in this case the two lengths satisfy
\begin{align}
    \bar L = \sqrt{1-\beta^2} L.
\end{align}
Our goal now is to compute the Poisson bracket of $L$ with $\bar L$ on the ADM phase space. But before even starting any calculation, one can already observe that this Poisson bracket is generically ill-defined. The reason being that the Poisson bracket between the canonical ADM variables, i.e.~the spatial metric and its conjugate momentum, is singular and involves a three dimensional Dirac delta distribution. This is nothing unusual in field theories, but it also means that in order to obtain a well-defined result of the Poisson bracket, one either attempts to perform a symplectic reduction of the system by restricting the spacetime manifold to the world sheet of the rod, or one has to consider quantities which provide a smearing of the bracket over a three dimensional spatial hypersurface. In the latter case, when looking at quantities such as the length of an object in some spatial direction, one must regularize this length by smearing it over the other two dimensions of space. This naturally introduces a regulator for the spatial directions orthogonal to the world sheet of the object under consideration. Namely, the expression for $L$ would become
\begin{align}
    L = \int_{R(t=0)} dx \sqrt{h(x)}\ \rightarrow\ L^{(\zeta)} = \int_{R(t=0)} dx^1 \int_{\mathbbm{R}^2} dx^2 dx^3 \varrho^{(\zeta)}(x^2, x^3) \sqrt{g_{11}(0,x^a)}\ 
\end{align}
such that $\varrho^{(\zeta)}(x^2, x^3)$ is a positive function, with a physical dimension of inverse length squared $[\text{Length}]^{-2}$, corresponding to a smooth regularization of the Dirac delta distribution parametrized by a parameter $\zeta > 0$. In other words, the function satisfies
\begin{equation}
    \int_{\mathbbm{R}^2} dx^2 dx^3 \varrho^{(\zeta)}(x^2, x^3) = 1
    \label{IntSmearingFunc}
\end{equation}
for any $\zeta$, and
\begin{align}
   \lim_{\zeta \rightarrow 0}\ \varrho^{(\zeta)}(x^2, x^3) = \delta(x^2) \delta(x^3).
   \label{LimSmearingFunc}
\end{align}
The regularization of one of the lengths suffices to smear the Poisson bracket, we then have
\begin{equation}
    \{L^{(\zeta)}, \bar L\} = \int_{R(t=0)} dx^1 \int_{\mathbbm{R}^2} dx^2 dx^3 \varrho^{(\zeta)}(x^2, x^3) \{g_{11}(0,x^a), \bar L\}.
    \label{PBlengths}
\end{equation}
We will further comment on this aspect of the regularization later in this section, once we have the final expression for the Poisson bracket $\{L^{(\zeta)}, \bar L\}$. 

Now, in order to evaluate the Poisson bracket $\{g_{11}(0,x^a), \bar L\}$ in \eqref{PBlengths},
one has to compute the Poisson bracket
\begin{equation}
    \{g_{11}(t, y), g_{11}(t',z) \}.
    \label{PBmetric}
\end{equation}
This means that this Poisson bracket generically involves computing variations at non-equal times. Indeed, given that the simultaneity surfaces of the two observers are not identical, since the observers are boosted with respect to each other, a more appropriate evaluation of the phase space commutation in this context may require a more covariant approach, i.e.~covariant brackets such as the Peierls brackets \cite{Peierls:1952cb, DeWitt:2003pm}. However, this can be avoided in the local setting we established earlier, by using an expansion of the metric tensor for the boosted observer around the point $p$, then computing the Poisson bracket on the ADM phase space order by order in the expansion. This approximation is the same we used to identify the simultaneity surfaces, and it will be sufficient to establish that the Poisson brackets \eqref{PBmetric} and \eqref{PBlengths} do not generally vanish. This is what we demonstrate in the following.

Given the expression of the induced metric $\bar h(x)$ in \eqref{rod.metric} and using the expression of the time variable $t$ in terms of $x$ as in \eqref{Sim.Surf.}, we can perform an expansion in $x$ of $\bar h$ around $x(p)=0$ and we get
\begin{align}\label{Exp.barh}
    \bar h(x) &= g_{11} + \left(\beta g_{11} \dot g_{11} + \partial_1 g_{11}\right) x - \beta^2 g_{11}^2 - \beta g_{11} \left(-\dot g_{11} + 2 \beta \Gamma_{111} + 2 \beta^2 g_{11} \dot g_{11} \right) x + {\cal O}(x^2) \notag \\[2ex]
    &= (1 - \beta^2 g_{11}) \bigl[ g_{11} + \left( 2\beta g_{11} \dot g_{11} + \partial_1 g_{11} \right) x \bigr]+ {\cal O}(x^2).
\end{align}

In the ADM formulation, the phase space variables are the spatial metric tensor $g_{ab}$ and its conjugate momentum $P^{ab}$ satisfying
\begin{equation}
    \{g_{ab}(y), P^{cd}(z)\} = \delta^c_{(a}\delta^d_{b)}\delta^{(3)}(y, z).
    \label{PBADM}
\end{equation}
We then have the relations
\begin{equation}
    \dot g_{ab} \coloneqq \partial_0 g_{ab} = 2NK_{ab} + ({\cal L}_{\vec N}g)_{ab}, \quad 
    P^{ab} = \frac{\sqrt{q}}{\kappa}\bigl(K^{ab} - g^{ab}K\bigr), \quad 
    K^{ab} = \frac{\kappa}{\sqrt{q}}\biggl(P^{ab} + g^{ab}\frac{P}{2}\biggr)
    \label{ADMexpressions}
\end{equation}
where $\kappa \coloneqq 16 \pi G$, $N$ is the lapse function, $\vec N = N^a$ is the shift vector, $K_{ab}$ is the spatial extrinsic curvature tensor and $K$ its trace, $q$ is the determinant of the spatial metric $g_{ab}$, while $P$ is the trace of the momentum $P^{ab}$.

Consequently we obtain
\begin{align}
    \bigl\{g_{ab}(y), K^{cd}(p)\bigr\} &= \frac{\kappa}{\sqrt{q(p)}}\Bigl\{g_{ab}(y), P^{cd}(p) + \frac{1}{2}g^{cd}(p) g_{ij}(p) P^{ij}(p)\Bigr\} \notag \\[2ex]
    &= \frac{\kappa}{\sqrt{q(p)}} \biggl(\delta^c_{(a}\delta^d_{b)} + \frac{1}{2}g^{cd}(p) g_{ab}(p) \biggr) \delta^{(3)}(y, p)
    \label{}
\end{align}
which in turn implies
\begin{align}
    \{g_{ab}(y), \dot g_{cd}(p) \} &= 2\kappa \frac{N(p)}{\sqrt{q(p)}} g_{ci}(p) g_{dj}(p) \biggl(\delta^i_{(a}\delta^j_{b)} + \frac{1}{2}g^{ij}(p)g_{ab}(p)\biggr) \delta^{(3)}(y, p) \notag \\[2ex]
    &= \frac{\kappa}{\sqrt{q}} \Bigl(g_{ac} g_{bd} + g_{ad} g_{bc} + g_{ab} g_{cd} \Bigr) \delta^{(3)}(y, p).
\end{align}
where we use the gauge we chose earlier (section 2), which fixes the lapse as $N=1$. Having this expression, we plug the result \eqref{Exp.barh} into \eqref{PBlengths} and we calculate the final expression of the Poisson bracket of the lengths:

\begin{align}
    \{L^{(\zeta)}, \bar L\} &= \int_{R(t=0)} dx^1 \int_{\mathbbm{R}^2} dx^2 dx^3 \varrho^{(\zeta)}(x^2, x^3) \int_{R(\bar t=0)} dy \left\{\sqrt{g_{11}(0,x^a)} , \sqrt{\bar h(y)} \right\} \notag \\[2ex]
     &= \int_{0}^{\epsilon} dx^1 \int_{\mathbbm{R}^2} dx^2 dx^3 \varrho^{(\zeta)}(x^2, x^3) \int_{0}^{\epsilon} dy \frac{\{g_{11}(0,x^a) , 2 \beta g_{11} \left(1 - \beta^2 g_{11} \right) \dot g_{11} \}}{4\sqrt{g_{11}(0,x^a)}\sqrt{g_{11}(1 - \beta^2 g_{11})}} y + {\cal O}(y^2) \notag \\[2ex]
    &= \frac{\beta}{4} \epsilon^2 \int_{0}^{\epsilon} dx^1 \int_{\mathbbm{R}^2} dx^2 dx^3 \varrho^{(\zeta)}(x^2, x^3) \sqrt{\frac{ g_{11} \left(1 - \beta^2 g_{11} \right)}{g_{11}(0,x^a)}} \{g_{11}(0,x^a) , \dot g_{11} \} + {\cal O}(\epsilon^3) \notag \\[2ex]
     &= \frac{3 \kappa \beta}{4} \epsilon^2 \int_{0}^{\epsilon} dx^1 \int_{\mathbbm{R}^2} dx^2 dx^3 \varrho^{(\zeta)}(x^2, x^3) \sqrt{\frac{ g_{11} \left(1 - \beta^2 g_{11} \right)}{g_{11}(0,x^a)}}\, \frac{g_{11}^2}{\sqrt{q}}\delta^{(3)}(x^a,0) + {\cal O}(\epsilon^3) \notag \\[2ex]
     &= \frac{3 \kappa \beta}{8} \frac{g_{11}^2\sqrt{1 - \beta^2 g_{11}}}{\sqrt{q}} \ell_\zeta^{-2} \epsilon^2 + {\cal O}(\epsilon^3)
    \label{PBlengthsExplicit}
\end{align}
where\footnote{The integration over $x^1$ with Dirac delta on the fourth line of \eqref{PBlengthsExplicit} is performed over a finite interval, with zero being at the boundary of the domain of integration. The result is generally ambiguous and depends on the prescription adopted to define the integration with Dirac delta. In our case, we adopt the prescription provided by a symmetric regularization of the Dirac delta, which implies that $\int_{0}^{\epsilon} dx\, \delta(x) f(x) = \frac{1}{2} f(0)$.} 
$\ell_\zeta^{-2} \coloneqq \varrho^{(\zeta)}(0, 0) > 0$.
The result in \eqref{PBlengthsExplicit} shows that the Poisson bracket of the two lengths (one being smeared) does not vanish, and consequently the lengths measured by two observers, boosted with respect to each other, do not generally commute. Also, the result depends on the relative velocity parameter $\beta$ and vanishes when $\beta=0$ as expected.
Additionally, one can see from the calculation in \eqref{PBlengthsExplicit} that the result does not depend on which length observable is regularized, whether it is for the observer at rest or the moving one, and failure to regularize one of the lengths would lead to an ill defined expression involving $\delta(0)$. This last aspect is still reflected in the parameter $\ell_\zeta$ which has a dimension of length, and as per \eqref{LimSmearingFunc} satisfies $\ell_\zeta \rightarrow 0$ as $\zeta \rightarrow 0$. As one might expect, the smearing introduces a fiducial length scale $\ell_\zeta$, i.e.~infrared scale, smoothing the physical length of the one dimensional rod over the remaining two orthogonal space dimensions.

Notice that in the case of Minkowski spacetime with metric $\eta = {\rm diag}(-1,1,1,1)$, the result in \eqref{PBlengthsExplicit} is exact and higher orders in $\epsilon$ would vanish. This is a consequence of the fact that higher orders in the expansion of the metric involve second derivatives or higher of the metric tensor, implying that the Poisson brackets of these higher orders will result into terms proportional to first derivatives and higher of the metric, which all vanish for the constant Minkowski metric. Hence, we get
\begin{align}
    \{L^{(\zeta)}, \bar L\} &= \frac{3 \kappa}{8} \beta \sqrt{1 - \beta^2}\, \ell_\zeta^{-2}\, \epsilon^2 
    \label{PBlengthsMinkowski}
\end{align}
which shows that even in Minkowski spacetime, geometric observables measured by observers boosted with respect to each other generically do not Poisson commute.

\section{Summary and discussion}

In summary, to answer the question of whether geometrical observables measured by different observers Poisson commute or not, we focused on computing the Poisson bracket of the lengths of a rod as measured by an observer at rest and a moving one. Using an expansion of the metric tensor in the coordinates, we derived an equation for the simultaneity surface of a given geodesic observer and employed it to compute the expressions of the lengths in different frames. It is worth emphasizing that the form of the simultaneity surface is in general rather non-trivial: Even for a stationary observer, the simultaneity surface generally does not coincide with the surface of constant time coordinate in the gauge $g_{00} = -1$, $g_{0a} = 0$. (For a more comprehensive discussion on this point, see e.g.~\cite{Lachieze-Rey:2001rrt, Bolos:2002qp}). We then calculated the Poisson brackets of the lengths seen by the two observers in the ADM formalism, leading to the non-trivial result in \eqref{PBlengthsExplicit}. Consequently, we established that geometrical observables measured in different frames do not generally Poisson commute. Furthermore, the result \eqref{PBlengthsMinkowski} shows that this conclusion remains valid even in Minkowski spacetime, which is an intriguing outcome that we further discuss below.

A very interesting observation is that if we were to consider the observers' world lines intersection point $p$ to be an arbitrary point of the rod, say such that $\epsilon_- \leqslant x \leqslant \epsilon_+$ with $\epsilon_-< 0 < \epsilon_+$, then the final result of the Poisson bracket of the two lengths becomes\footnote{Note that the result in \eqref{PBlengthsExplicit} cannot be obtained from the one in \eqref{PBlengthsGen} simply by taking the limit $\epsilon_- \to 0$; in order to get the correct numerical prefactor, one has to take into account the factor of $1/2$ which arises when the Dirac delta is integrated over a segment of the positive half-line starting from the point $x = 0$.}
\begin{align}
    \{L^{(\zeta)}, \bar L\}
     &= \frac{3 \kappa \beta}{4} \frac{g_{11}^2\sqrt{1 - \beta^2 g_{11}}}{\sqrt q} \ell_\zeta^{-2} (\epsilon_+^2 - \epsilon_-^2) + {\cal O}(\epsilon^3).
    \label{PBlengthsGen}
\end{align}
If the rod were to be symmetric with respect to the point $p$ (i.e.~$\epsilon_- = - \epsilon_+$) then the Poisson bracket \eqref{PBlengthsGen} vanishes at second order in $\epsilon$. This is in fact a particular case where all even orders of the expansion in $\epsilon$ would vanish because of the symmetric interval of integration. However, this does not mean that the total Poisson bracket vanishes, because the higher orders ($>2$) of odd power in $\epsilon$ would not vanish and the total Poisson bracket of the lengths remains non-trivial. Generally, this aspect is simply an artifact of the choice of coordinates. Yet, if the spacetime is Minkowski and the point $p$ is indeed the proper midpoint of the rod, then the Poisson bracket in \eqref{PBlengthsMinkowski} would actually vanish. However, in light of the discussion in section 2 and from the perspective of observers meeting at the midpoint, the Poisson bracket of the lengths of the rod corresponds to the sum of two independent contributions canceling each other, each contribution being given by the Poisson bracket of the lengths of one half of the rod. One can track the sign difference to be sourced in the time ordering. Indeed, looking at figure \ref{fig:Measurement}, one can deduce that for one half of the rod, say the one on the left, the simultaneity surface of the inertial observer lies in the future of the simultaneity surface of the moving observer, and vice versa for the second half of the rod. In our calculations, this sign difference manifests through the sign of the difference between the coordinate time of $p$ and that of a point on the rod in the coordinates expansion.

Let us now discuss further the result in \eqref{PBlengthsMinkowski}. If we take the smearing function $\varrho^{(\zeta)}$ to be sharply peaked around $(x^2,x^3)=(0,0)$, and assume that the width of the peak is of the order ${\cal O}(\ell_\zeta)$, then at lowest order in $\epsilon$ and $\ell_\zeta$, the lengths $L^{(\zeta)}$ and $\bar L$ are given by
\begin{equation}
	L^{(\zeta)} \simeq \sqrt{g_{11}}\, \epsilon \qquad \text{and} \qquad \bar L \simeq \sqrt{g_{11}(1 - \beta^2 g_{11})}\, \epsilon.
	\label{}
\end{equation}
This allows us to write the result \eqref{PBlengthsExplicit} approximately in terms of the lengths as
\begin{align}
    \{L^{(\zeta)}, \bar L\} \simeq \frac{3 \kappa \beta}{8} \ell_\zeta^{-2}\frac{g_{11}}{\sqrt{q}} \,  L^{(\zeta)}\, \bar L.
    \label{GeneralRes}
\end{align}
In Minkowski spacetime, the Cartesian coordinate length $\epsilon$ becomes a physical length, \mbox{i.e.~$L = \epsilon$} and $\bar L = \sqrt{1 - \beta^2} \, \epsilon$. Also, thanks to the property in \eqref{IntSmearingFunc}, the regularized length $L^{(\zeta)}$ coincides exactly with the physical length $L$. Hence, in the special case of flat spacetime, we obtain the exact result
\begin{align}
    \{L^{(\zeta)}, \bar L\} &= \frac{3 \kappa \beta}{8}\, \ell_\zeta^{-2}\,  L\, \bar L.
    \label{MinkowskiRes}
\end{align}
Upon quantization of the gravitational degrees of freedom, the Poisson bracket presumptively becomes the commutator divided by $i\hbar$. Therefore, on grounds of the result \eqref{GeneralRes}, which holds in an arbitrary spacetime, we would expect the commutator of the length observables to be given by an expression of the form
\begin{align}
    [L^{(\zeta)}, \bar L] &\simeq 6\pi i\, \beta\, \frac{\ell_P^2}{\ell_\zeta^2}\, \cO_g \, L^{(\zeta)}\, \bar L 
    \label{ComRes}
\end{align}
where $\ell_P \equiv \sqrt{\hbar G}$ is the Planck length, and $\cO_g$ is some gravitational operator representing the quantization of the factor $g_{11}/\sqrt{q}$ in eq.~\eqref{GeneralRes}. We then see that the commutator of the lengths, as measured in two relatively moving frames would depend on the relative velocity parameter $\beta$, and would vanish when $\beta=0$ as expected. Furthermore, the commutator depends on both length observables and the quantum gravity scale, i.e.~the Planck length.

It is interesting to compare the expression \eqref{ComRes} with the computation of the analogous commutator of areas performed in \cite{Rovelli:2002vp}. The result obtained in that work for the Poisson bracket $\{A, \bar A\}$ is proportional to the time derivative of the area and therefore vanishes in Minkowski spacetime. This apparent contradiction is explained by the fact that in \cite{Rovelli:2002vp}, the intersection point of the two observers' world lines is fixed from the beginning to coincide with the geometric center (the two-dimensional counterpart of the midpoint of the rod) of the rectangular surface under consideration. With this setup for the calculation, the contribution from the terms of linear order in the expansion \eqref{Ind.metric} is identically vanishing, and it is necessary to take the expansion of the induced metric to the second order in the coordinates in order to obtain a non-trivial result. However, as shown by the calculation presented in section \ref{sec:lengths}, this does not mean that the commutator generically vanishes in Minkowski spacetime; rather, it is merely a particular case in which two fundamentally independent contributions cancel each other in the final result. Our approach provides a more general treatment and reveals that the non-commutativity holds generically even in flat spacetime.

Consequently, the refutation of the paradox raised by Amelino-Camelia \cite{Amelino-Camelia:2002cfs} and by Magueijo and Smolin \cite{Magueijo:2001cr}, presented in \cite{Rovelli:2002vp}, appears ineffective in this context, since the paradox itself is formulated precisely in Minkowski spacetime, where the Poisson bracket is assumed to vanish. In our work, we prove that the Poisson bracket \eqref{MinkowskiRes} does not vanish even for Minkowski spacetime and the argument concerning the apparent incompatibility of the minimal length (area, volume, ...) with Lorentz symmetry is systematically resolved, thanks to the fact that the geometric observables measured by two boosted observers do not commute and cannot be simultaneously measured in a quantum theory of gravity.

\acknowledgments
\vspace{-12pt}

The authors would like to thank Simone Speziale for fruitful discussions.
J.KG. and L. V. acknowledge the support from the European COST Actions BridgeQG CA23130 and CaLISTA CA21109. 
For MA this research was funded by the National Science Centre, Poland, through SONATA~19 grant no.~2023/51/D/ST2/00296. For IM this work was funded by National Science Centre, Poland through grant no.~2022\slash 44\slash C\slash ST2\slash 00023.
For the purpose of open access, the authors have applied a CC BY 4.0 public copyright license to any author accepted manuscript (AAM) version arising from this submission. 

\vspace{-12pt}

\bibliography{References}
\bibliographystyle{uiuchept}

\end{document}